\documentclass[twocolumn]{revtex4}
\usepackage{latexsym}
\usepackage{graphicx}
\usepackage{units}
\usepackage{bm}

\begin{document}

\title{Production of a chromium Bose-Einstein condensate}
%\subtitle{}
\author{Axel Griesmaier,
        J\"urgen Stuhler,
        Tilman Pfau}
\affiliation{5. Physikalisches Institut,
            Universit\"at Stuttgart,
            Pfaffenwaldring 57,
            70550 Stuttgart,
            Germany\\
            \email{a.griesmaier@physik.uni-stuttgart.de}
            }
%\email{a.griesmaier@physik.uni-stuttgart.de}
%\date{Received: date / Revised version: date}

\begin{abstract}
The recent achievement of Bose-Einstein condensation of chromium
atoms~\cite{Griesmaier:2005a} has opened longed-for experimental
access to a degenerate quantum gas with long-range and anisotropic
interaction. Due to the large magnetic moment of chromium atoms of
\unit[6]{$\mu$B}, in contrast to other Bose-Einstein condensates
(BECs), magnetic dipole-dipole interaction plays an important role
in a chromium BEC. Many new physical properties of degenerate gases
arising from these magnetic forces have been predicted in the past
and can now be studied experimentally. Besides these phenomena, the
large dipole moment leads to a breakdown of standard methods for the
creation of a chromium BEC. Cooling and trapping methods had to be
adapted to the special electronic structure of chromium to reach the
regime of quantum degeneracy. Some of them apply generally to gases
with large dipolar forces. We present here a detailed discussion of
the experimental techniques which are used to create a chromium BEC
and allow us to produce pure condensates with up to $10^5$ atoms in
an optical dipole trap. We also describe the methods used to
determine the trapping parameters.
\end{abstract}
\maketitle
\section{Introduction}
\label{intro} Among the species that have been used to create
degenerate quantum gases, chromium is outstanding with respect to
its very large magnetic dipole moment of \unit[6]{$\mu_B$} (where
$\mu_B$ is Bohr's magneton) in the ground state $^7S_3$. This dipole
moment leads to a magnetic dipole-dipole interaction (MDDI) which
has a strength comparable to the interaction arising from the
isotropic and effectively short range interatomic (contact)
potential. Since the MDDI scales with the square of the magnetic
moment, it is a factor of 36 stronger for chromium than for alkali
atoms. Although atomic Bose-Einstein condensates are very dilute
systems, strength, range and symmetry of the weak interactions
between the atoms determine the essential properties of degenerate
quantum gases and give rise to many phenomena that can be
observed.\\
In the Bose-Einstein condensates that have been produced so far
\cite{Anderson:1995a,Davis:1995,Bradley:1995,Fried:1998,Modugno:2001,Robert:2001,Weber:2003a,Takasu:2003},
interaction between the atoms is governed by the contact potential.
Many exciting effects of these interactions have been studied (for
an overview see e.g. \cite{Pitaevskii:2003,Varenna99}) and are
subject to still growing interest of both, theorists and
experimentalists. Recently, BECs with contact interaction have been
used as model systems for solid-state physics problems like the
Mott-metal-insulator transition~\cite{Greiner:2002a,Stoferle:2004}.
Tuneable contact interactions have been used to realize new states
of quantum matter like a Tonks-Girardeau gas~\cite{Paredes:2004} and
to produce molecular Bose-Einstein
condensates~\cite{Duerr:2004,Herbig:2003}. The cross\-over from a
molecular BEC to a degenerate Fermi gas has been
studied~\cite{Greiner:2004,Bartenstein:2004,Zwierlein:2004,Bourdel:2004}
and superfluidity and vortices have been observed in a strongly
interacting Fermi gas~\cite{Zwierlein:2005}.
In contrast, the MDDI between chromium atoms is long-range and
anisotropic. The presence of an interaction with different symmetry
and range leads to new properties of such a gas. A change in the
aspect ratio of the condensate after ballistic expansion of a
dipolar condensate has been predicted~\cite{Giovanazzi:2003a} and
could be observed very recently~\cite{Stuhler:2005}. If one of the
Feshbach-resonances of chromium~\cite{Werner:2005} is used to tune
the contact-interaction close to zero, the MDDI can even become the
dominant interaction. For such a regime, new kinds of quantum phase
transitions ~\cite{Yi:2004} have been predicted. The ground state
and stability~\cite{ODell:2004a,Goral:2002,Santos:2000a} of a
dipolar BEC can also be studied. In a pancake-like trapping
geometry, the occurence of a Maxon-Roton in the excitation spectrum
has been predicted~\cite{Santos_Roton03}. Furthermore the MDDI is
also tuneable~\cite{Giovanazzi:2002a} which allows for studies of a
degenerate quantum gas in different regimes where atoms are
interacting dominantly
via either short-range and isotropic or long-range and anisotropic potentials.\\
\section{Bose-Einstein condensation of ${^{52}}$Cr}

\label{sec:BEC}
\begin{figure}
\includegraphics[width=1\columnwidth]{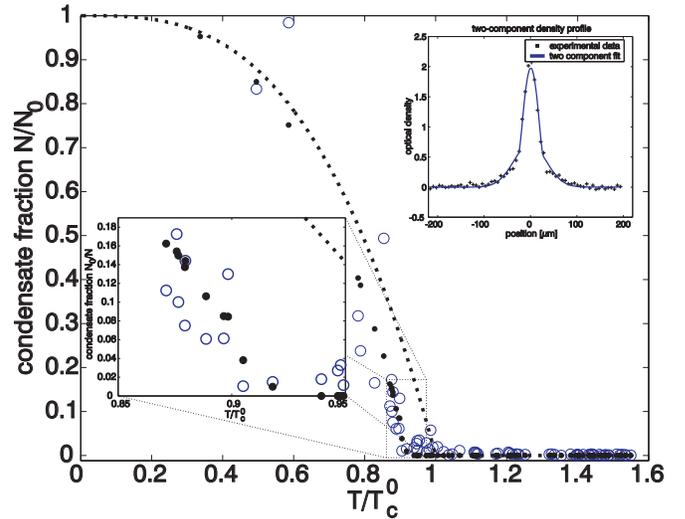}
\caption{ \label{fig:formation} Condensate fraction ($N_0/N$)
dependence on temperature relative to the transition temperature of
an ideal gas ($T/T_C^{0}$),
${T_C^{0}\approx0.94\frac{\hbar\omega}{k_B}N^{1/3}}$. Open circles
represent the measured data. Black dots represent the predicted
fraction ${\frac{N}{N_0}=1-{\left(\frac{T}{T_C}\right)}^{3}}$ where
$T_C=T_C^0+\delta T_C^{int}+\delta T_C^{fs}$. ${\delta
T_C^{fs}=-0.73\frac{\overline{\omega}}{\omega}N^{-1{}/3}T_C^0}$ is a
shift in the critical temperature due to the finite number of atoms
and ${\delta T_C^{int}=-1.33\frac{a}{a_{HO}}N^{1{}/6}T_C^{0}}$ takes
into account the contact interaction~\cite{Giorgini:96}. Here
$a=105a_0$ is the chromium scattering length~\cite{Werner:2005},
$a_0$ being Bohr's radius, $a_{HO}$ is the harmonic oscillator
length, $T$ is the temperature of the thermal cloud, ${\omega}$ is
the geometric and ${\overline{\omega}}$ the arithmetic mean of the
trap frequencies. The dashed curve shows the dependence for the
ideal gas.}
\end{figure}

Bose-Einstein condensation of chromium is achieved in a crossed
optical dipole trap using the methods described in
section~\ref{sec:production} of this paper where
\begin{figure}
\includegraphics[width=1\columnwidth]{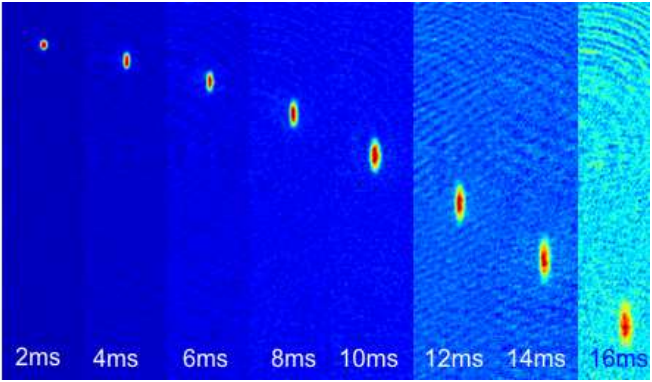}
\caption{ \label{fig:expand} Time of flight absorption images of an
expanded almost pure condensate of about 40.000 atoms after
expansion times between \unit[2]{ms} and \unit[16ms]. The pictures
are in the y-z plane of our setup, z being the horizontal
direction.}
\end{figure}
a schematic illustration of the setup can also be found. We identify
the presence of a Bose-Einstein condensate by anisotropic expansion
when the condensate is released from an anisotropic trap with a
tighter confinement in x- and y-direction and a less tight
confinement z-direction. Figure \ref{fig:expand} shows time of
flight images of such a condensate taken after variable time of
ballistic expansion between \unit[2]{ms} and \unit[16]{ms}. The
appearance of a two-component distribution in the density profile of
the cloud  also signalizes the presence of a condensed fraction. The
point of emerging degeneracy becomes obvious when the number of
atoms in the condensate and the thermal fraction are determined
separately for different final laser powers. These numbers can be
obtained by fitting a two-dimensional two-component distribution
function to the density profiles of the cloud. The inset of
fig.~\ref{fig:formation}) shows a cut through the center of such a
distribution with a line displaying the result of a two-component
fit. If the trapping potential is well known, not only the number of
atoms in the thermal cloud but also the temperature of the cloud can
be calculated from the width of the Gaussian. We have precisely
measured the trap frequencies in all directions and at different
laser powers using the methods described in section~\ref{sec:freq}.
This calibration allows us to calculate temperatures, densities and
phase space densities from one time-of-flight picture. The
temperatures obtained with this method have been compared to the
results of much more precise time-of-flight series and are
concordant with these results within a maximum deviation of $8\%$ if
the time of flight is not too short. The determination of the number
of atoms and widths vastly depends on the quality of the images. To
reduce fringe patterns usually present in absorption images due to
fluctuations of laser power and intensity or mechanical vibrations
between the separate images, we have implemented a method developed
by the Sengstock group~\cite{Brinkmann:2005a} in our image
evaluation software. This method allows the later reduction of
fringes without loss of information.
This way, we where able to improve our previous
analysis~\cite{Griesmaier:2005a} of the dependence of the fraction
of condensed atoms $(N_0/N)$ on the ratio of the temperature of the
cloud to the critical temperature $(T/T_C)$. The corresponding data
are shown as open circles in fig.~\ref{fig:formation}. When we
approach the critical temperature from above $(T/T_C>1)$, a kink in
the condensate fraction plot marks the onset of Bose-Einstein
condensation and provides an experimental value for the critical
temperature of $T_{exp}\sim $\unit[700]{nK}. Based on the trap
frequencies, number of atoms and temperature, we have also
calculated the expected condensate fraction when finite size effects
as well as a correction arising from the contact
interaction~\cite{Giorgini:96} are taken into account. These
expected values are represented by black dots in
figure~\ref{fig:formation} and demonstrate a
very good agreement of our data with the predicted dependence.\\
\section{Producing a chromium BEC}
\label{sec:production}
\begin{figure}
\includegraphics[width=1\columnwidth]{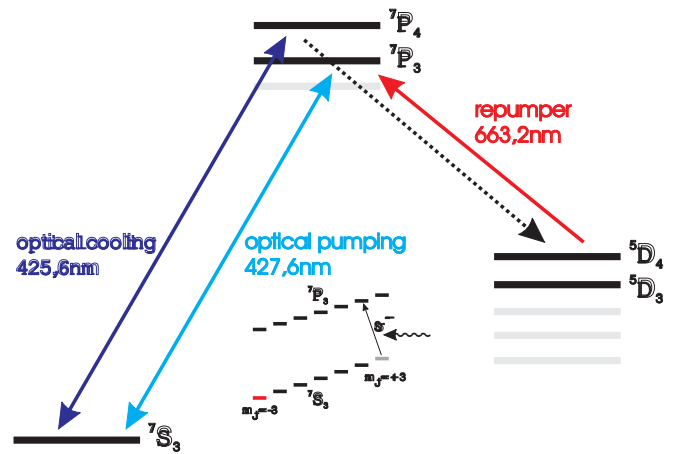}
\caption{ \label{fig:levels} Chromium levels and lines relevant for
cooling and optical pumping.}
\end{figure}
\begin{figure}
\includegraphics[width=1\columnwidth]{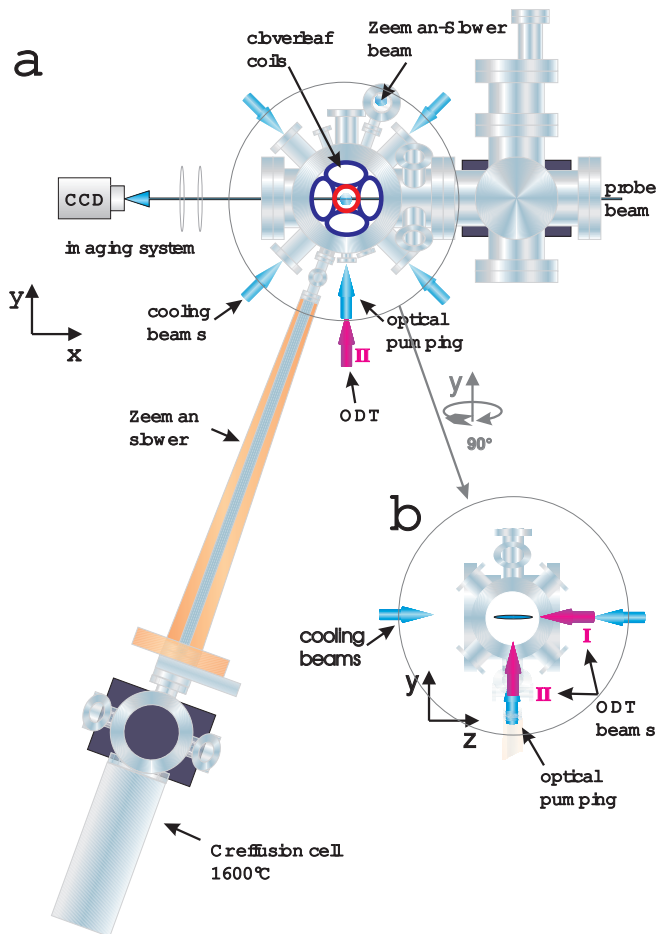}
\caption{ \label{fig:chamber} Schematic setup of our experiment: a)
whole apparatus, b) upper chamber seen in the direction of the probe
beam. The horizontal beam I of the dipole trap propagates in
z-direction and carries a power of up to \unit[9.8]{W}. The vertical
beam II defines our y-direction and has a maximum power of
\unit[4.8]{W}.}
\end{figure}
\begin{figure}
\includegraphics[width=1\columnwidth]{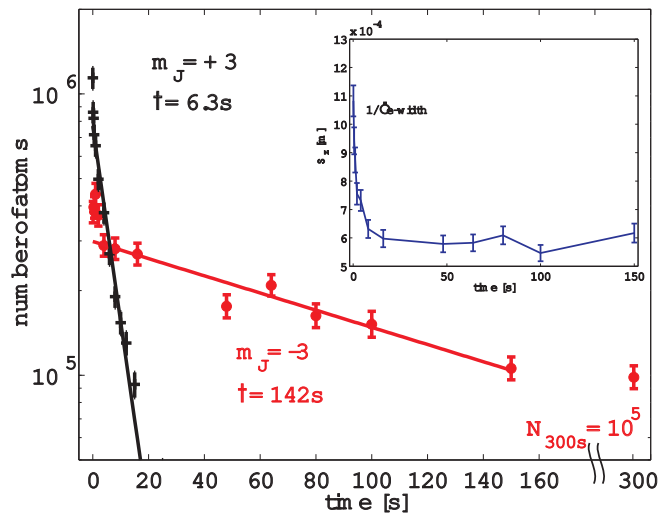}
\caption{ \label{fig:lifetime} Comparison of trap lifetimes before
(crosses) and after (circles) pumping the atoms to the lowest Zeeman
substate. The inset shows the change of the axial size of the
expanded cloud in time of flight after different holding times.}
\end{figure}

The unique electronic structure of chromium (configuration $[{\rm
Ar}]3d^54s^1$) with 6 unpaired electrons with aligned spins in the
$^7S_3$ ground state gives rise to a total spin quantum number of 3
and a very large magnetic moment of \unit[6]{$\mu_B$} making it a
well suited element for magnetic trapping. It also has a versatile
level-scheme allowing optical cooling and pumping techniques. A
drawback however is an extremely large dipolar relaxation rate also
arising from the large magnetic moment. In magnetic trapping
potentials, trapped atoms are necessarily in the so called low-field
seeking states ($m_J=+3$) and therefore have a high Zeeman energy.
Dipolar relaxation processes can lead to a redistribution over the
$m_J$ levels. The released Zeeman energy leads to atom loss and
heating. In earlier experiments~\cite{Hensler:2003a,Hensler:2003b},
these relaxation processes prevented chromium from condensation in a
magnetic trap because as the atoms are cooled, the growing spatial
density leads to higher and higher two-body loss and heating rates.
To circumvent this loss, one has to polarize the atoms in the
energetically lowest Zeeman sub state $m_J=-3$ where the atoms can
not gain any more energy by flipping their spin but would rather
have to bring up energy~\footnote{This effect has recently been
suggested to be used for a novel cooling
scheme~\cite{Hensler:2005}.}. Therefore dipolar relaxation is
suppressed by energy conservation. This state, however, is not
magnetically tappable and optical trapping techniques have to be
used, where the trapping potential is independent of the Zeeman
state~\cite{Grimm:2000a}. The step to an optical dipole trap (ODT)
and polarization of the atoms in $m_J=-3$ together with an exact
alignment of the trapping beams are the essential steps in the
preparation process of a chromium
BEC which will be described in detail in this section.\\
Figure~\ref{fig:chamber} illustrates schematically the setup which
is used for our experiments. The preparation scheme starts with the
generation of a beam of chromium atoms by a high temperature
effusion cell operating at \unit[1600]{$^o$C} and subsequent
deceleration by a Zeeman slower. Chromium has a $\Lambda$-like level
scheme with weekly allowed intercombination transitions from the
excited state of the optical cooling cycle
${^7}S_3{}\leftrightarrow{}{^7}P_4$ at \unit[425.6]{nm} into long
lived metastable $^5D_3$ and $^5D_4$ states (branching ratio
$250000:1$). The relevant levels and transitions are shown in
fig.~\ref{fig:levels}. Chromium has a high magnetic moment in the
ground state as well as in the metastable states. This allows for a
continuous loading scheme of optically cooled atoms directly into a
magnetic trap with Ioffe-Pritchard (cloverleaf) configuration
(CLIP-trap)~\cite{Schmidt:2003a,Stuhler:2001}. With this technique
we are able to accumulate about $1.3\cdot10^8$ atoms within
\unit[10]{s} at roughly the Doppler temperature of
\unit[124]{${\mu}$K} and a phase space density of a few times
$10^{-9}$ in the metastable $^5D_4$ state. After the steady state
number of atoms in the magnetic trap is reached, the cooling lasers
are switched off and the atoms in the $^5D_4$ state are pumped back
to the ground state using a diode laser system resonant with the
${^5}D_4\leftrightarrow{}{^7}P_3$ transition at \unit[663.2]{nm}.
The magnetic trap is then beeing compressed fully by ramping up
currents through the trap coils. In doing so the cloud is heated up
again to $\sim$ \unit[1]{mK} and Doppler cooling of the optically
dense cloud~\cite{Schmidt:2003c} is performed within the trap at an
offset field of \unit[14]{G} using the cooling beams propagating in
z-direction. Doppler cooling increases the phase space density by
two orders of magnitude without losing atoms. Subsequently, the
currents through the coils are lowered to form a trap with an aspect
ratio as close as possible to the later shape of the optical dipole
trap to provide for efficient transfer between the traps. The offset
field in the center of the magnetic trap is adjusted close to
\unit[0]{G} by applying some extra current in the bias coils. In
this trap, we perform radio frequency (rf) evaporation with an
rf-sweep composed of $3$ linear rf-ramps from \unit[45]{MHz} down to
\unit[1.25]{MHz}. After this step, the cloud contains $4\cdot10^6$
atoms at a phase space density of $10^{-5}$ and a temperature of
\unit[22]{$\mu$K}. The overall gain in phase space density during
the magnetic trapping phase is thus about four orders of magnitude
while the number of atoms is
reduced by $1.5$ orders of magnitude.\\
Beginning from the first step of the rf-ramp, a single beam ODT in
horizontal direction with the same symmetry axis as the magnetic
trap is shone in. This trap is formed by a beam (marked as I in
fig.~\ref{fig:chamber} b) with a maximum of \unit[9.3]{W} at a
wavelength of \unit[1064]{nm} produced by a \unit[20]{W} Yb-fibre
laser (IPG PYL-20M-LP). The beam is focussed to a waist of
\unit[30]{$\mu$m} in the center of the magnetic trap. The linear
polarization of the light allows to split up the light into two
beams with adjustable ratio using a polarizing beam splitter and a
$\lambda/2$-plate. The second beam carries up to \unit[4.5]{W} and
is in a later stage shone in in vertical direction (marked as II in
fig.~\ref{fig:chamber}) to form a crossed dipole trap with the first
beam. Both beam intensities can be controlled
independently by acousto-optical modulators.\\
The final frequency of the rf-sequence is chosen such that the
highest number of atoms is remaining in the single beam trap after
ramping down the magnetic trapping potential within \unit[100]{ms}.
Perfect alignment of the ODT with best possible overlap between
magnetic and optical trap is essential for an efficient transfer
into the ODT. In particular, the first beam must be aligned in
perfectly horizontal direction because of its rather poor
confinement in axial direction resulting in a longitudinal trap
frequency of $\sim$ \unit[13]{Hz}. In radial direction, the trap
frequency is $\sim$ \unit[1450]{Hz}. Best alignment is achieved by
taking absorption images of the dipole trap within the magnetic trap
from the side and from above and looking at the alignment of the
optical trap with respect to the magnetically trapped cloud. The
horizontal beam is walked by using the final two mirrors in the
optical path to match direction and position of the beam with the
magnetic trap. The longitudinal position of the focus can be varied
by moving the final lens which is mounted on a translation stage
such that after switching off the
magnetic trap, the optical trap stays at the same position.\\
Immediately after the magnetic trap is off, we pump the atoms to the
lowest Zeeman state using about \unit[0.5]{mW} of \unit[427.6]{nm},
$\sigma{}^-$ polarized light and an intensity on the order of
\unit[1]{mW/cm$^2$}, resonant with the
${^7}S_3\leftrightarrow{}{^7}P_3$ transition for which $m_J=-3$ is a
dark state. The light is produced by a frequency doubled
master-slave diode laser system and shone in for \unit[1]{ms}. To
have a well defined alignment of the spins, the pumping process is
performed at an offset field of \unit[9]{G} in y-direction. In
contrast to commonly used rf techniques, this optical pumping
technique has the advantage that the occupation numbers of the
Zeeman substates are not only inverted but the polarization is also
purified. This is necessary in our experiment because in the
magnetic trap all low-field seeking states are occupied due to
dipolar relaxation. The efficiency of the transfer is close to
$100\%$ which is reflected in a dramatic increase of the lifetime of
the trapped cloud from $\tau_{\rm+3}\sim$\unit[6]{s} in $m_J=+3$ to
$\tau_{\rm-3}>$\unit[140]{s} in $m_J=-3$ (see.
fig.~\ref{fig:lifetime}). The magnetic offset field used for pumping
is kept on during all further preparation steps to prevent
thermal redistribution of the spins.\\
\begin{figure}
\includegraphics[width=1\columnwidth]{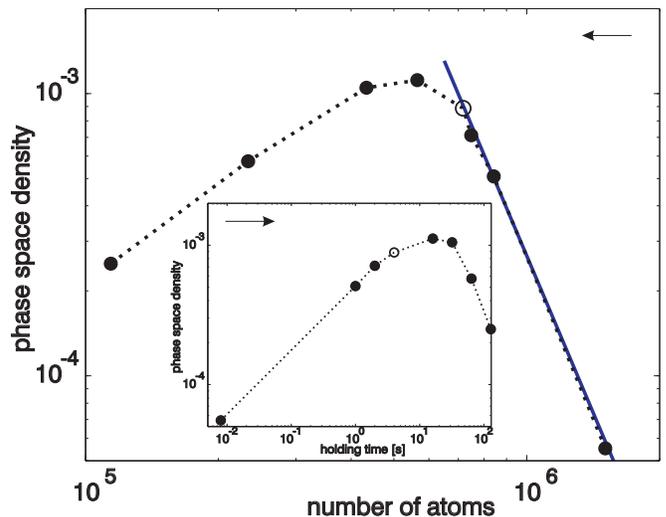}
\caption{ \label{fig:psdsingle} Double logarithmic plots of the
phase space density vs. number of remaining atoms and vs. holding
time in the single beam trap (inset). Arrows mark the chronology in
the plots. An open circle marks in both plots the optimum point in
time to start forced evaporation.}
\end{figure}
The transfer efficiency from the magnetic to the optical trap is
$40\%$ and we start with $1.5\cdot10^6$ atoms in the ODT. They have
an initial temperature of \unit[60]{$\mu$K} which corresponds to
about half the depth of the single beam trap (\unit[130]{$\mu$K}).
The phase space density is $5\cdot10^{-5}$, a factor of five higher
than the phase space density found when looking at the expansion of
the cloud after the final rf-ramp without an optical trap present,
although the temperature in the optical trap is significantly
higher. During the first seconds in the ODT, a very fast decay is
observed which becomes slower after about half of the atoms are lost
(fig.~\ref{fig:lifetime}). Along with this decrease of the number
comes a very large increase in the phase space density which was
determined from time of flight pictures taken during the lifetime of
the trap. We therefore attribute the observed loss to essentially
pure plain evaporation. Figure~\ref{fig:psdsingle} shows the
evolution of the phase space density during the first \unit[120]{s}
in the ODT plotted vs. the number of remaining atoms in the main
plot and vs. time in the inset. The straight line in the main graph
has a slope of $3.6$ orders of magnitude gain in phase space density
per lost order of magnitude in the number of atoms. This illustrates
the very high efficiency of the plain evaporation. We let this
evaporation proceed for \unit[5]{s} until it gets less efficient
which is identifiable by the data points snapping off from the straight line.\\
\begin{figure}
\includegraphics[width=1\columnwidth]{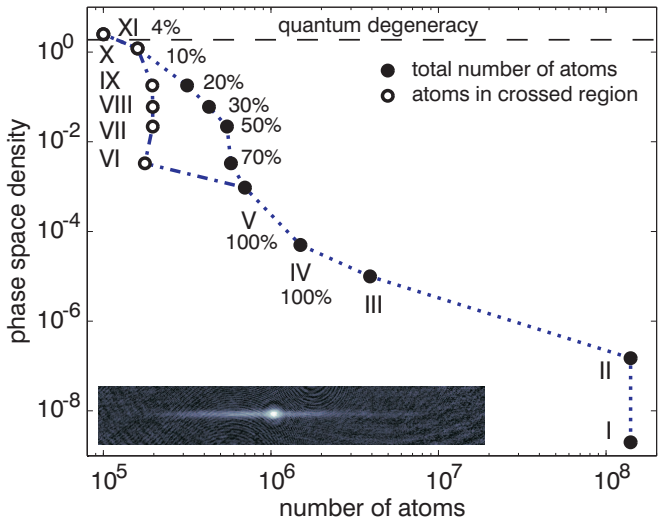}
\caption{ \label{fig:psdmap} Evolution of phase space density vs.
number of atoms during the whole preparation process starting with
$1.3\cdot10^8$ atoms after the CLIP-loading procedure (I). The next
steps are Doppler-cooling (II), RF-cooling (III), transfer into the
ODT (IV) and initial plain evaporation in the ODT (V). Evaporation
is forced from (VI) to (XI). From step (VI) on, filled circles
represent the total number of atoms while open circles represent
only the atoms trapped in the steep potential in the crossed region.
For steps (IV) to (XI), the power in the horizontal trapping beam in
percent of the maximum power is also displayed. The vertical beam is
ramped up between steps (IV) and (V). The inset shows a typical
image of a cloud at $70\%$ power in the horizontal beam, where a
large number of atoms are still located in the wings of the
horizontal trapping beam.}
\end{figure}
During plain evaporation, the second dipole trap beam with a waist
\unit[50]{$\mu$m} ((II) in fig.~\ref{fig:chamber} b) is ramped up to
full power to form a crossed dipole trap with high trap frequencies
in all directions. Best alignment of this beam is achieved by
ramping down the first beam, taking an absorption image and
improving from shot to shot the number of trapped atoms by adjusting
the position of the second beam's focus. Finally, the positioning is
adjusted to the number of condensed atoms. In the cooling process,
we proceed with forced evaporation in the crossed trap by ramping
down the power of the horizontal beam in $6$ steps within
\unit[5]{s} to $70\%$, \unit[1.8]{s} to $50\%$, \unit[1.1]{s} to
$30\%$, \unit[1.2]{s} to $20\%$, and \unit[1.6]{s} to $10\%$ which
is just above condensation. In fig.~\ref{fig:psdmap} the phase space
densities after each of these evaporation steps are represented by
steps (VI) to (XI). From step (VI) on, open circles represent the
number of atoms trapped in the crossed region of the trap.
Particularly during the first evaporation steps, a larger number of
atoms are still trapped in the outer regions ("wings") of the first
beam shown in the inset of figure~\ref{fig:psdmap}. During the
evaporation process, atoms are permanently loaded from the wings
into the crossed region, causing the number of atoms in this region
to stay almost constant during steps (VI) to (IX). After stage (IX)
finally all remaining atoms are in the crossed trap and the wings of
the horizontal beam are empty. From there on we lose less than half
of the atoms in steps (X) and (XI) until quantum degeneracy is
reached at a remaining $\sim8\%$ of the initial power corresponding
to $\sim$ \unit[800]{mW}. Large condensates with up to $100.000$
atoms are created when we ramp down to $4\%$ of the maximum power.
This last ramp takes \unit[250]{ms} and we additionally hold the
trap at constant power for another \unit[250]{ms} to let the system
equilibrate before we
release the atoms from the trap to take an absorption image.\\
The overall gain in phase space density during the optical trapping
period starting at $5\cdot10^{-5}$ with $1.5\cdot10^6$ atoms and
ending on the order of $1$ with $10^5$ atoms is more than $4$ orders
of magnitude, while we lose only $1.2$ orders in the number of
atoms. Thus the overall quotient of orders of PSD gain vs. orders of
atom loss is $>3.6$, almost twice as large as in the magnetic trap.
This suggests the use of higher laser power for the optical trap to
form a deeper potential and therefore be able to reduce the time of
rf-evaporation in the magnetic trap and transfer the atoms into the
ODT already at an earlier stage.
\section{Trap frequency measurement}
\label{sec:methods} \label{sec:freq}
\begin{figure}
\includegraphics[width=1\columnwidth]{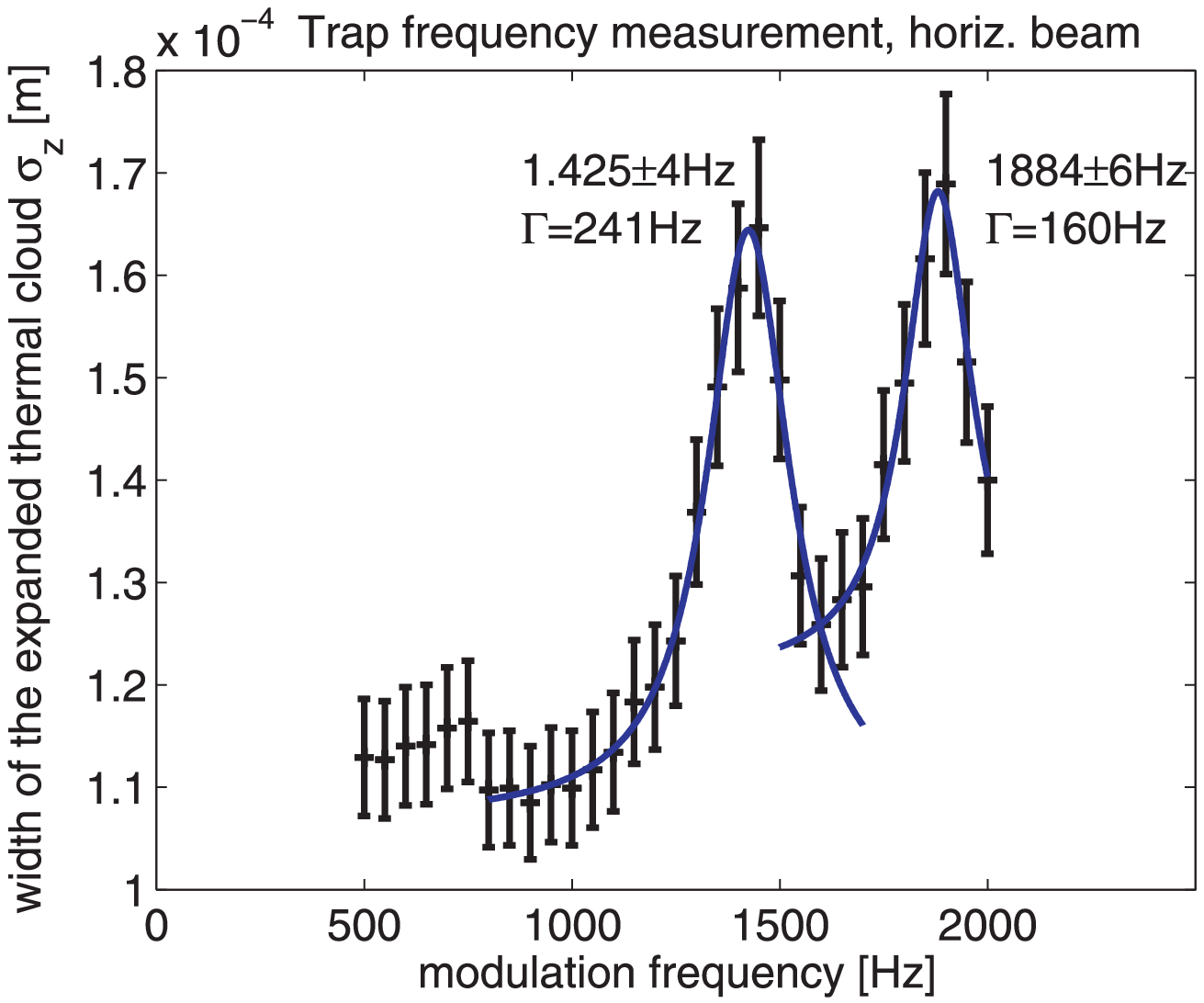}
\caption{ \label{fig:trapfreqh} Measurement of the trap frequencies
by modulation of the horizontal beam. $1/\sqrt{e}$-width of the
cloud in z-direction after time of flight plotted over the
modulation frequency. The pronounced peaks at \unit[1425]{Hz} and
\unit[1884]{Hz} are at twice the trap frequencies in y- and
x-direction, respectively.}
%\end{figure}
%\begin{figure}
\vspace{1cm}
\includegraphics[width=1\columnwidth]{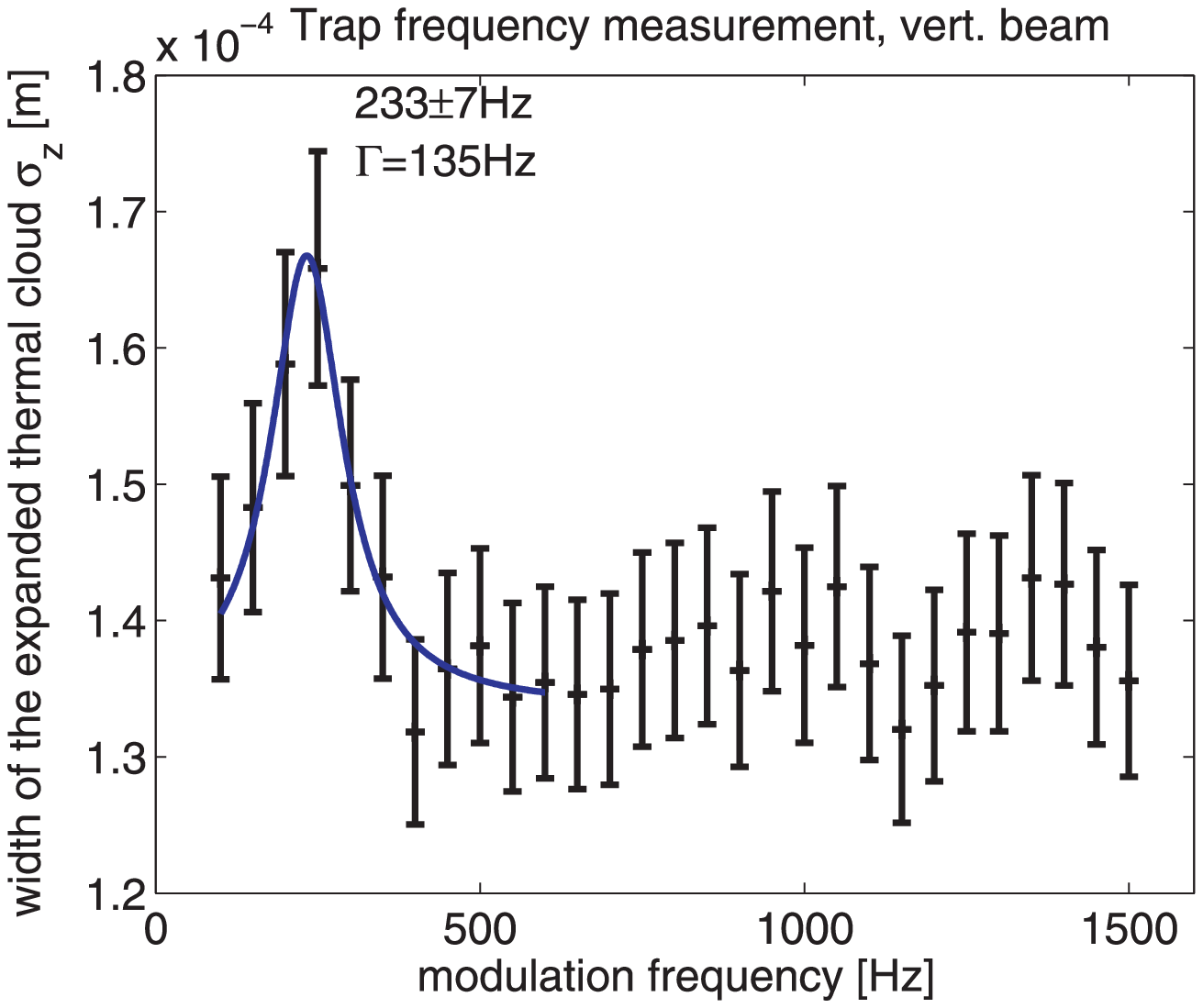}
\caption{ \label{fig:trapfreqv} Measurement of the trap frequencies
by modulation of the vertical  beam. The resonance peak is at
\unit[233]{Hz}, twice the trap frequency in z-direction.}
\end{figure}
Exact knowledge of the trap frequencies is necessary for all
measurements where the density distribution of the atoms in the trap
is important and where the phase space density of the sample has to
be known. In the crossed dipole trap, we measure them independently
using a parametric heating technique with a trapped thermal cloud.
The preparation of the sample is the same as for the condensation
except that the evaporation ramp is stopped before reaching
degeneracy. The intensities of the trapping beams are then
adiabatically (within $\sim$\unit[100]{ms}) changed to form the
trapping potential that has to be calibrated. After the cloud is
prepared in this way, a remote programmable function generator
(Stanford Research Systems DS345) in burst mode is used to modulate
the intensity of one of the trapping beams by a few percent ($<5\%$)
by varying the rf-power of one of the AOMs sinusoidally for
$\sim$\unit[500]{ms}. In a series of experiments, the preparation
scheme is then repeated, varying from shot to shot the frequency at
which the function generator modulates the laser power. After each
experiment, an absorption image of the cloud is taken after a time
of free ballistic expansion of \unit[4]{ms}. If the modulation
frequency is close to twice the trap frequency in one direction
${\omega_{\rm mod}=2\omega_{\rm trap}}$ or subharmonics
${\omega_{\rm mod}=2\omega_{\rm trap}/n}$, energy is transferred to
the atoms and the atomic cloud heats up~\cite{Grimm:2000a}. This
heating effect is most effective at twice the trap frequency which
is reflected in a decrease of the number of
atoms and a drastic increase of the width of the expanded cloud.\\
In a crossed dipole trap, each beam mainly contributes to the trap
frequencies only in the directions orthogonal to the beam, whereas
its contribution in the longitudinal direction is rather small and
this frequency is mainly determined by the other beam. Therefore
modulating the intensity of one beam allows to resolve only the
frequencies in the two orthogonal directions. To measure the trap
frequencies in all three directions, the described sequence has to
be performed twice, modulating the horizontal beam in one sequence
and the vertical beam in the other. Because the temperature and
therefore the size of the expanded cloud shows much less shot to
shot fluctuations than the number of atoms, we plot the measured
widths of the cloud for the whole series over the used modulation
frequencies and fit a Lorentz-function to the data around each
resonance peak. Examples of such measurements are shown in
figures~\ref{fig:trapfreqh}~and~\ref{fig:trapfreqv}.
\section{Conclusion}
In conclusion, we have described in detail the improved methods used
to generate Bose-Einstein condensates of chromium atoms. With these
methods, we are able to produce condensates with $10^5$ atoms which
is a very good basis for promising further experiments. The
preparation makes use of magneto-optical, magnetic and optical
techniques which are adapted to the special electronic and magnetic
properties of chromium. The evolution of phase space density during
the preparation process has been discussed and the methods used to
characterize the optical trap have been presented.
\section*{Acknowledgments: }
We thank all members of our atom optics group for their
encouragement and practical help. We thank Luis Santos, Paolo Pedri,
Stefano Giovanazzi, and Andrea Simoni for stimulating discussions.
This work
was supported by the SPP1116 and SFB/TR21 of the German Science Foundation (DFG).\\

\label{sec:conclusion}

\end{document}